# The automatic detection of lumber anatomy in epidural injections for ultrasound guidance

**Farhad Piri,[a] Sima Sobhiyeh,[b] Amir H. Rezaie,[a] Faramarz Mosaffa,[c]**

[a] Amirkabir University of Technology, Electrical Engineering Department, Hafez Avenue, Tehran, Iran,15875-4413.
[b] Louisiana State University, School of Electrical Engineering and Computer Science, Baton Rouge,LA, USA,70803.
[c] Shaheed Beheshti University of Medical Sciences, Akhtar hospital, Department of Anesthesiology, Elahieh St., Tehran, Iran, 1955841452.

**Abstract**. The purpose of this paper is to help the anesthesiologist to find the epidural depth automatically to make the first attempt to enter the path of the needle into the patient's body while it is clogged with bone and avoid causing a puncture in the surrounding areas of the patient`s back. In this regard, a morphology-based bone enhancement and detection followed by a Ramer-Douglas-Peucker algorithm and Hough transform is proposed. The proposed algorithm is tested on synthetic and real ultrasound images of laminar bone, and the results are compared with the template matching based Ligamentum Flavum (LF) detection method. Results indicate that the proposed method can faster detect the diagonal shape of the laminar bone and its corresponding epidural depth. Furthermore, the proposed method is reliable enough providing anesthesiologists with real-time information while an epidural needle insertion is performed. It has to be noted that using the ultrasound images is to help anesthesiologists to perform the blind injection, and due to quite a lot of errors occurred in ultrasound-imaging-based methods, these methods can not completely replace the tissue pressure-based method. And in the end, when the needle is injected into the area (dura space) measurements can only be trusted to the extent of tissue resistance. Despite the fairly limited amount of training data available in this study, a significant improvement of the segmentation speed of lumbar bones and epidural depth in ultrasound scans with a rational accuracy compared to the LF-based detection method was found.

**Address all correspondence to:** Farhad Piri, Amirkabir University of Technology, Electrical Engineering Department, Hafez Avenue, Tehran, Iran, 15875-4413; Tel: +98 921-440-2097; E-mail: farhad.peeri@aut.ac.ir

## 1 Introduction

An epidural injection is a technique whereby a local anesthetic drug is injected via a catheter placed into the epidural space providing temporary or prolonged relief from pain for patients during childbirth or after surgery[31], however, administering needle insertion for catheter placement of lumbar epidural anesthesia is a challenging procedure[1,2,3,19], and since the correct needle depth can vary due to individual patients, overshoot in depth measurement can cause spinal nerve injury.[1,3,4] Currently, the only available method for epidural injection is a technique called the blind method in which the physician thrusts his hand blindly into the tissue to decide the epidural injection site.

When the needle is in the Ligamentum flavum, the resistance reaches plateau and rapidly drops sensibly as the needle tip reaches the epidural space. This intuitive technique is named loss-of-resistance (LOR) which is applied to determine the needle tip position.

In LOR, there is no display to assist the doctor which obviously will cause multiple needle attempts at needle placement besides pain and discomfort to the patient. Because epidural injection occurs around the spinal cord, the faults in injection may lead to paralysis of the patient.[4]

Predicting the accuracy of needle placement before puncture is impossible with any of the landmark-based techniques.[13]

Recently, Ultrasound (US) imaging is becoming the clinical standard to assist forward peripheral nerve blocks.[14–16] Moreover, US imaging is non-invasive, safe, and simple to use technology, it also provides real-time images, and there is no radiation threat.[13]

The use of the US in epidural anesthesia helps to reduce complications, multiple needle pass, visual analog scale, the rate of side effects, and increase accuracy and rigor in this procedure.[3,13,17,18]

Nevertheless, US-guided epidural injection has some drawbacks too. For example, ultrasound does not provide clear visibility, especially due to the thin and deep structure of the epidural space and the contamination induced by the speckle noise.[19]

Ultrasound is used for aiding epidural needle insertion, but differentiating spinal structures due to noise, artifacts, and inexperience of anesthesiologists are matters of interpretation. Moreover, preserving sterile conditions while anesthesiologist requires to measure relative distances has great importance; thus, interaction with the ultrasound devices must be limited. Also, there are time limitations in the operating room. Thus to conclude, automated measurement has a great deal of help.[3]

With the use of automatic detection of lumbar anatomy[3] for Ultrasound-guided epidural injections, the presence of the imaging physician in the operating room is no longer needed.

An automatic method needed for detecting laminae and LF is useful for helping the anesthesiologist quickly choose the appropriate paramedian plane and measure the skin-to-LF depth which represents the required needle insertion depth from the skin surface, as minimal interaction with the ultrasound machine is required. [3]

If the information is automatically extracted from ultrasound images, we will take advantage of high accuracy as well as many other benefits including the following:

-Maintaining sterile conditions during surgery; physician anesthesia`s contact with Ultrasonic devices can be minimized.

- Just a short train will be provided for the anesthesiologist to work with the ultrasound machine, and he is no longer responsible for the analysis of the ultrasound images.

- Refraining from adding an ultrasound physician to the medical team necessary for the surgery of the patient.

Ultrasound imaging of the lumbar spine is performed by employing three ultrasound planes: paramedian, median transverse, and median longitudinal.[9,13]

Studies demonstrated that the paramedian plane is the best window to see the epidural space and neuraxial structures compared with the median transverse or median longitudinal planes, therefore the transverse plane could not be a successor for the measurement of the epidural space depth, although others have found success with it.[19] Recently, automatic detection algorithms of lumbar bone in transverse and midline view with the help of machine learning techniques have been studied.

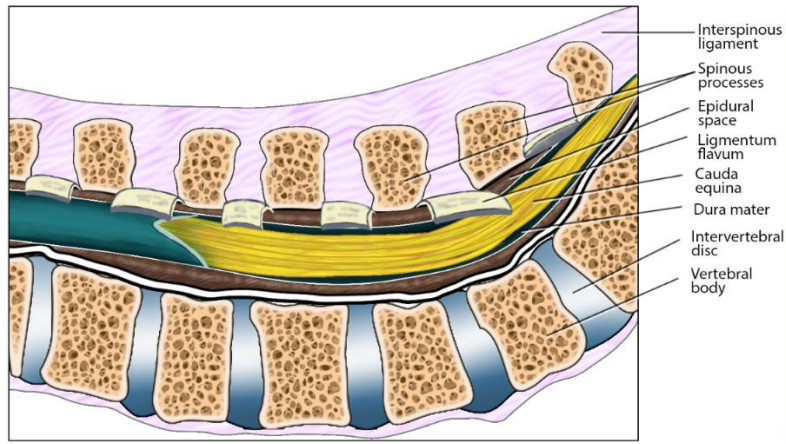

**Fig .1** Lumbar anatomy of an ultrasound image.

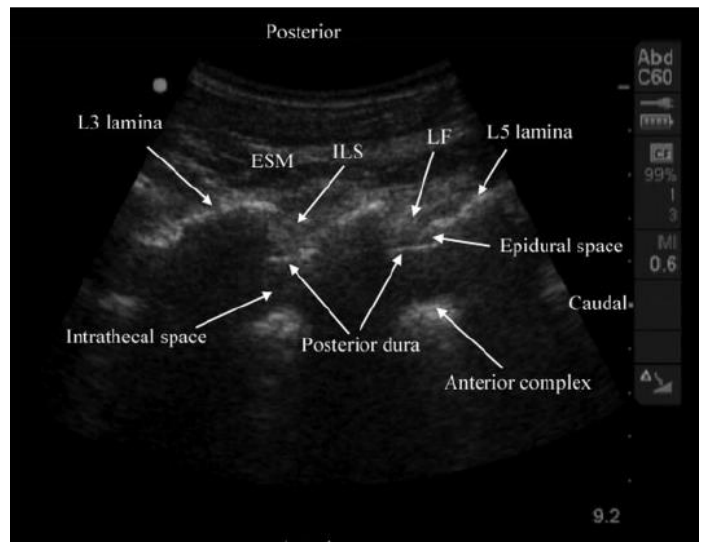

**Fig .2** Paramedian sagittal plane in the ultrasound image of the lumbar anatomy.[43]

Yu et al. (2013) [32] proposed an automatic method to detect the needle insertion site in ultrasound images of epidural anesthesia in the transverse plane. The method employed normalized difference of Gaussians to remove speckle noise and applied a method of template matching of signatory features dominant in the transverse view combined with a position correlator to detect the bone structures of the inter-spinous images. The method was extended by utilizing several classifiers: a cascading classifier[33], a neural network[34], and a support vector machine[35]. Florian Berton et al.

(2016)[36] extracted several image features from the ultrasound spine image and used a linear discriminant analysis classifier on a large data set for the vertebrae segmentation. The previously cited methods have analyzed the ultrasound images in the transverse plane. In our case, we utilize the paramedian plane, where the preamble window is larger and the identification of Ligamentum flavum, dura mater and cauda equine nerves is easily approved.[9]

Kerby et al. (2008)[37] used panorama images of lumbar levels to create an image with peaks and valleys onto which 3 filters were applied: a median filter to bring out the bone surfaces properly and two linear filters that operated in vertical and horizontal directions. The vertical and horizontal filters were applied to the image to enhance bone edges and the periodic nature of the vertebrae, respectively. Later, the local maxima of the filtered image were selected as the corresponding vertebrae level. In a similar work, Al-Deen Ashab et al. (2013)[38] used this technique to overlay the lumbar levels onto the back of the patient and obtained an error within a clinically accepted range. Two latter techniques were adopted in the paramedian plane using panoramic ultrasound images which is not the case in our single-image-based analysis. Tran, D., & Rohling, R. N. (2010)[3] proposed the first automatic detection of lumbar anatomy in ultrasound images of the paramedian plane. The method adopted in their paper is elaborated in what follows.

First, this method applies spatial compounding with warping on the ultrasound images from different angles and combines them using the adaptive median to improve image quality. Next, the phase symmetry algorithm is applied to the resulted image to enhance bone (lamina) and Ligamentum flavum (LF) ridges. Then, a synthetic lamina template is matched to this ridge map using Pearson's cross-correlation, and the most likely lamina locations are found. Finally, the LF template is traversed along the lower region of the lamina, and the highly correlated positions of the LF is obtained.

Incorrect alignment of frames can result in blurring of spatially compounded images. This can stem from the variation of sound speed by as much as 14% in soft tissues introducing distortions due to refraction and speed-of-sound (SoS) errors.[24,25] An additional non-rigid registration (warping) was previously proposed to re-align the features of images.[23] Therefore, warping which takes place according to the Pearson cross-correlation as the similarity measure is used to improve visibility of the Ligamentum flavum and the epidural space.

For feature extraction, phase symmetry is used which can be calculated by

$$PS(x) = \frac{\sum \lfloor [|e_n(x)| - |o_n(x)|] - T \rfloor}{\sum_n \sqrt{e_n(x)^2 + o_{n(x)}^2 + \varepsilon}}, \quad (1)$$

Where $\varepsilon$ is a small number added to prevent division by zero and $T$ is a threshold used to offset the phase symmetry by the expected noise value. The Log–Gabor filter is used to extract the even and odd parts to be used in the phase measures. Where $e_n(x)$ is the even part of the signal and $o_n(x)$ is the odd part of the signal produced by the Log–Gabor filter.

As the features mostly appeared at specific orientations (near oblique to the ultrasound beam), only three orientations ($90°$, $120°$ and $150°$) were used.

The expected ridge map is produced by multiplying the image intensity with phase symmetry matrix. The next step is the extraction of the lamina from the ridge map which is done by matching the template of the lamina to the ridge map. Lamina template is a diagonal line of angle matching the average angles of the lamina blurred by a Gaussian filter. Also, diagonal angular Gaussian blurred filter is blurred at an angle of 90° to the direction of its length by a thickness blur to widen the capture range. These two filters are convolved together to result in the lamina template.

This template is matched to the image using the Pearson correlation coefficient measure[4,27], and a lamina similarity map is generated. Correlations in the lamina similarity are used to choose the location of the most likely lamina. Since adult subjects in that study had at most three laminas in the ultrasound image, the highest similarity point in the image was chosen as the location of the first lamina, then, the search region for the second lamina is set to at least 20 mm away from the first lamina location, as the distance between two laminas is approximately 30 mm.[28] After finishing the repeated procedure of finding lamina bones, they are eliminated from the ridge map. For LF detection a different template is used. The cross-correlation $R_{LF}$ of the LF template with the ridge map, is only calculated for pixels around the lamina. A subtle way to do this is to look for the point with the largest difference between $R_{LF}$ and $R_{\text{lam}}$ to emphasize the end of the ridge which is defined as

$$LF_n(x,y) = \text{argmax}_{i,j}(R_{LF}(i,j) - R_{\text{lam}}(i,j)), \tag{2}$$

where $LF_n(x,y)$ is the location of the *n*th LF.

Considering imaging methods of the lumbar region, ultrasound uses the pulse-echo technique to create images. If reflection from bone is strong enough, it casts shadows in the beam direction. Artifacts that may affect image quality are reverberation, refraction and tissue type such as, excessive fat in obesity which generally is considered as a noise present throughout the image. Spatial compounding[21] uses beam steering, to capture several pictures of the same region from different angles of incidence. Speckle noise is dependent on the distribution of reflectors along the ultrasound path[22]. Since speckle pattern decorrelates with beam angle, beam steering will lead to different speckle noise patterns. Averaging images with varying patterns of noise but similar anatomical features will reduce the angle-dependent artifact, such as noise and enhance the LF structure of interest.

The work that is done in this paper is a big step and the first proposal to automate the lumbar anatomy detection process, although it has some defects that we discuss some of them below;

1. In the approach taken in the LF detection algorithm, to improve the quality of ultrasound images, the image compounding method has been used which requires the ultrasound machine to conduct multi-angle imaging and unify the right combination of images from different angles. However, compound imaging devices are expensive and not always available. Therefore, in general, it is required to use the methods of improving the quality of ultrasound images.

2. In the aforementioned approach, the template matching method has been used to identify the tissues in the image which is an excellent candidate to find the exact bone structures; however, since this method detects the middle of each tissue and does not find the edges around the tissue, it still suffers from the error rate of a few millimeters because the anesthesiologist tears the inner layer of yellow ligament and then reaches the injection space. The depth of needle insertion reported by the anesthesiologist is the space between the inner layer of the yellow ligament and the skin`s surface not the space between the yellow ligament and the skin's surface. It goes unnoticed that the running time of the template matching algorithm is high, and as a result, the template matching approach is not suitable for real-time applications without utilizing parallel computing or expensive hardware.

This paper aims to detect the epidural space depth in medical ultrasound images to offer the anesthesiologist real-time visual aid in finding a bone-free path for needle insertions such as the epidural anesthesia needle insertion. Since it is difficult especially for the inexperienced anesthesiologist to find a bone-free way without a lot of struggling, it is of great importance to offer a visual help for them to make ultrasound guidance screening tool more efficient.[9]

In a recent study, Sobhiyeh S. et al. (2014)[18] proposed a despeckling algorithm to satisfy a machine doing segmentation. However, due to the inexperience of anesthesiologists in ultrasound image interpretation and to hold sterile conditions by decreasing the interaction of anesthesiologists, the despeckling algorithm was performed followed by a new algorithm firstly proposed for lamina bone detection. In other words, the bone detection algorithm is a post-processing step which follows the despeckling action. Hence, the proposed bone detection algorithm is evaluated by its comprehensiveness in different images and execution time which has a significant role in choosing the most efficient algorithm to perform ultrasound image processing techniques. In a recent study, the LF-based automatic detection algorithm utilizing phased symmetry and template matching approaches had high accuracy in finding LF but it suffered a time consumption inefficacy which is improved in our morphology-based method.

This method is performed on a synthetic image which is created in Photoshop$^{TM}$ software and then simulated in Field II$^{TM}$ software as an image to cover the same noise pattern existent in ultrasound imaging. Because of the nature of the Hough transform and its modus operandi, we believe that it is the right candidate for bone detection since it is fast enough in finding the positions of bones of interest and their volunteer orientations.

Moreover, few methods have been studied to see if the combination of Hough transform, phase symmetry, and morphology techniques can result in a more efficient approach.

The purpose of this paper is to help the anesthesiologist to find a way to make the first attempt to enter the path of the needle into the patient's body while it is clogged with bone and avoid causing a puncture in the surrounding areas of the patient`s back. It has to be noted that using the ultrasound images is to help anesthesiologist to perform the blind injection, and due to quite a lot of errors occurred in ultrasound imaging-based methods, these methods can not completely replace the tissue pressure-based method, even though, when the needle is passed into the epidural space, measurements can only be trusted to the extent of the tissue resistance.

The organization of this paper is as follows: in Sec. 2, the complex wavelet diffusion (CWD) for ultrasound image speckle-noise reduction is presented. The morphology-based epidural space detection approach followed by a Hough transform is analyzed in Sec. 3, and the experimental results regarding real and synthetic ultrasound images of the laminar bone are discussed. Later in Sec. 4, Clinical trials and methods along with a fair evaluation of the proposed algorithm are presented. Finally, the conclusion of the paper is given in Sec. 5.

## 2   The objective of Noise Reduction and Segmentation

Ultrasound images substantially suffer from the speckle noise which makes the refined structures hard to distinguish from the surrounding background and reduces accuracy in tissue detection and medical diagnosis[39,18]. Hence, prior to lumbar anatomy detection in ultrasound images, a valuable noise reduction solution is demanded. As mentioned earlier, Sobhiyeh S. et al. (2014)[18] proposed a noise reduction method called Complex Wavelet Diffusion (CWD) to better highlight the laminas and remove the speckle noise. The CWD method operates in various directions, enhancing the visibility of diagonally oriented laminar bones while removing speckle noise. Therefore, the CWD method has been used as a pre-processing step for our lumbar anatomy detection approach, which utilizes the enhanced laminar bone to better identify the epidural space depth.

*2.1 Noise Reduction & Edge Enhancement (Image preprocessing)*

Speckle removal techniques are aimed at offering an enhancement in the visibility of ultrasound images and improvement in the performance of automatic segmentation. Generally, most speckle filters are designed to increase the visibility of ultrasound images while preserving the texture information. However, for segmentation purposes, image texture does not assist in separating the objects by their boundaries. In this sense, removing the image texture is better to be considered. In addition, texture removal results in a more comprehensible image, which considerably improves the performance in object detection through an increase in speed and accuracy of the automatic segmentation. It goes unnoticed that an image simplification through piecewise smoothing of image regions in which edges are drawn clearly, regularly results in segmentation improvement. Mumford shah has explained the "cartoon" model which illustrates such simplification in terms of mathematical functions.[41]

To address the image simplification and denoising, non-linear anisotropic diffusion methods that generate a successively more and more blurred image in each iteration can be well-thought-out. Sobhiyeh et al. (2014) proposed a non-linear anisotropic diffusion (AD) method based on complex wavelet filter banks to remove speckle noise and enhance laminar bone visibility. Complex wavelet filter banks operate in six directions: $\pm 15, \pm 45, \pm 75$. Therefore, CWD performs better in enhancing the laminar bone which appears as a diagonal structure in the paramedian plane of ultrasound images (see Fig. 1.). The directional form of the AD equation in the Fourier domain is defined as

$$\hat{\tilde{f}}(u,v) = A_s \cdot A_a \cdot \tilde{f}(u,v) + \hat{D}_{1\,s} \cdot \{\hat{c}_1 * [\hat{D}_{1\,a} \cdot \hat{f}(u,v)]\} + \hat{D}_{2\,s} \cdot \{\hat{c}_2 * [\hat{D}_{2\,a} \cdot \hat{f}(u,v)]\} + \hat{D}_{3\,s} \cdot \{\hat{c}_3 * [\hat{D}_{3\,a} \cdot \hat{f}(u,v)]\} + \hat{D}_{4\,s} \cdot \{\hat{c}_4 * [\hat{D}_{4\,a} \cdot \hat{f}(u,v)]\} + \cdots, \qquad (3)$$

where $\hat{f}(u,v)$ represents the noisy image, $\hat{\tilde{f}}(u,v)$ represents the denoised image after a single iteration, $A$ denotes the low-pass filters known as approximation filters, $D_i$ denotes the high-pass filters called detail filters in direction $i$, and the indices $a, s$ indicate the discrimination between the analysis and synthesis filter banks. To discriminate between the speckle noise and the lumbar structures, arithmetic mean of the normalized detail coefficients in a random homogeneous region is chosen as a threshold. The diffusion function in the CWD method looks for the homogeneous regions and applies the diffusion function in the Speckle Reduction Anisotropic Diffusion (SRAD) method since this function smoothens the speckle noise. Whereas, for regions with higher edge information, Weickert filter[42] is applied since this filter is a better candidate for maintaining strong edges in an image, and $\rho$ is the calculated diffusion coefficient defined as

$$\rho_i(u,v) = \begin{cases} 0, & oth. \\ 1.5 \times \exp\left\{\frac{-[\eta_i^2(u,v) - \lambda_i^2]}{\lambda_i^2(1+\lambda_i^2)}\right\}, & 0 < \eta_i(u,v) \leq \lambda_i \\ 1.8 \times \exp\left\{\frac{-3.315}{\left[\frac{\eta_i(u,v)}{\lambda_i}\right]^4}\right\}, & \eta_i(u,v) > \lambda_i \end{cases}, \quad (4)$$

where $\kappa_i$ is the calculated edge detection variable specified by the size of the window $s$, $\mu_s$ is the arithmetic mean of the sliding window s, the operator $mean\{\}$ is the statistical mean of all the coefficients in the homogeneous region, and $\lambda_i$ is the constant value holding the information about the homogeneous region in each direction, scale, and iteration

$$\eta_i(u,v) = Normalized\{D_i(u,v)\} = \frac{D_i(u,v)}{\mu_s}, \quad (5)$$

$$\lambda_i = \begin{cases} \alpha \times mean\{Normalized\{D_i^h(u,v)\}\}, & j = 1 \\ \frac{\alpha \times mean\{Normalized\{D_i^h(u,v)\}\}}{\sqrt{2}^j}, & j \geq 2 \end{cases}. \quad (6)$$

Note that, in scales more significant than one, the detail coefficients tend to be smoother. Hence, the threshold value is divided to preserve edge information.

## 3 Lamina Crest Extraction

Section 3 presents the proposed algorithm for the detection of lamina bone and its corresponding epidural depth in steps and analyzes a phase-symmetry-based alternative method.

*3.1 Overview of the Method*

An overview of the proposed algorithm for epidural space depth detection is depicted in Fig. 3.

**Fig. 3** Overview of the lamina crest extraction algorithm

The first step in our algorithm is to remove spackle noise while improving lamina crest visibility through the CWD speckle removal method. Next, two morphological dilations are applied to the denoised image and then binarized by Otsu`s thresholding method. Then, the edge map is generated by applying a simple gradient operator to provide the linear information for the Hough transform procedure. With the edge map of bone structures in hand, the Hough transform is applied to detect the linear segments in the image. After locating the linear segments in a fixed range of orientations covering the lamina crest angle, the lowermost segment and its lower endpoint are chosen as the lamina bone of interest and the epidural space depth, respectively. Each step is elaborated in what follows.

   A. *Image preprocessing.*

Ultrasound speckle regularly reduces lamina bone visibility. The objective of noise reduction is to increase lamina uniformity while keeping the overall shape of the structures in the image. Complex wavelet diffusion (CWD) is a suitable image pre-processing technique to remove speckle noise in our lamina extraction based application. To begin with, the directional nature of wavelet filter banks results in better enhancement of laminas which appear as oriented parallel structures in paramedian view of epidural space. Moreover, the ringing effect caused by the wavelet filter bank helps to visualize the lamina shape properly[18]. Therefore, the CWD denoising method which is optimized for lamina visibility enhancement and speckle reduction is exploited to prepare the raw input for forthcoming image processing techniques. The difference between the denoised image and the raw input is shown in Fig. 4.

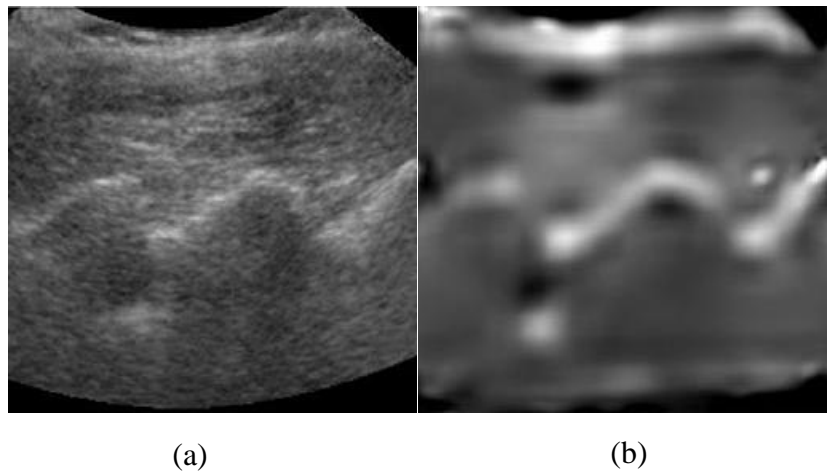

(a)            (b)

**Fig. 4** Result of the CWD denoising method: (a) raw input (b) denoised image

The following values are used as a basis for all of our data set in the CWD denoising method. The value of alpha is equal to 16, the number of iterations in anisotropic diffusion is set to 3, the number of wavelet transform scales is set to 7 and the diffusion coefficient used is the same as Eq.4.

B. *Morphological operations and thresholding*

Morphological operators are employed to point out the edge of the lamina as it curves in to form the lower part of the lamina. Also, as the dilation operators grow the size of the laminas, the thresholding step can better distinguish the bone structures.

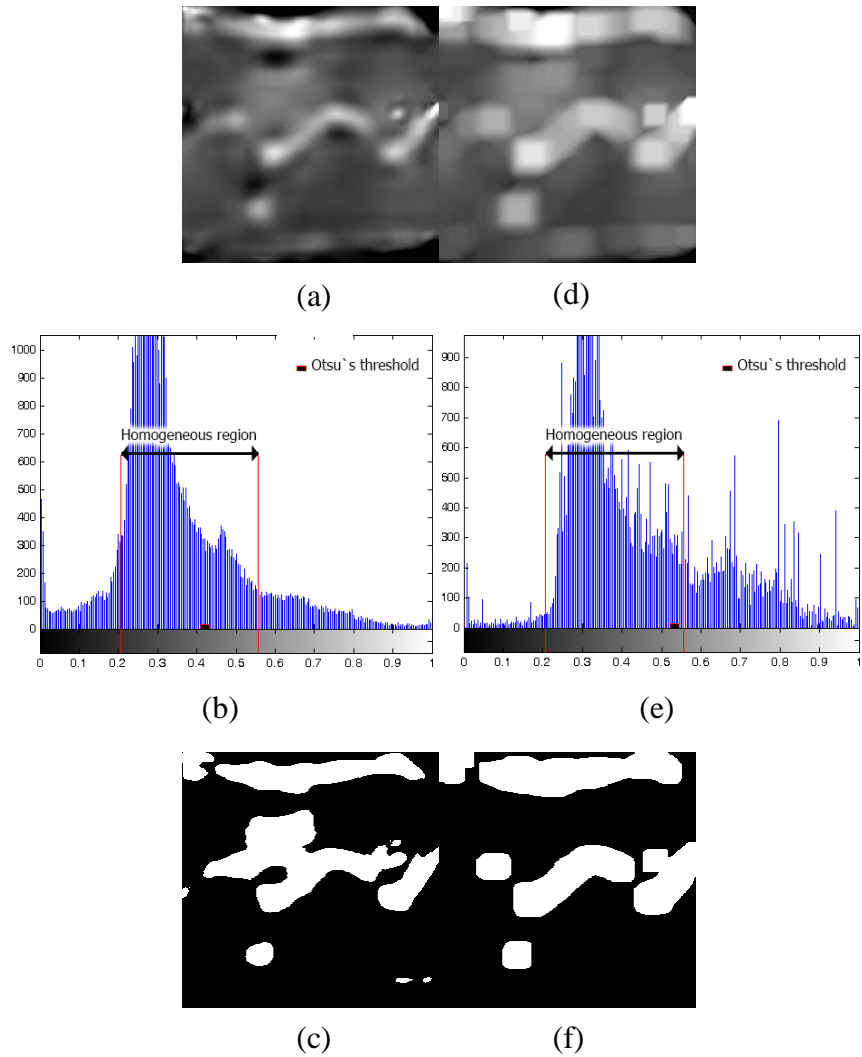

**Fig. 5** Results of Otsu`s thresholding method on: (a) original denoised image, and (c) double dilated image and its histogram which shows that dilations help the Otsu`s method differentiate between the homogeneous region and the bony structures.

As shown in Fig. 5, dilation operators help the thresholding method to better distinguish the bone structures as the dilation operation grows the continuous bone structures.

Fig. 5 shows the histograms of the denoised image both before and after applying morphological dilation operators. Therefore, three steps are taken into account: First, a dilation operation with a square-shaped structuring element is applied to the image. Then, the same step is repeated to form an even more obvious square shape in the lower end of the lamina. Finally, an erosion operation with a disk-shaped structuring element, but with a bigger size than the previous one, is applied to the resulted image to provide an image with a squared shape at the lower end of the lamina. Emphasizing the point at the lower part where the lamina curves in is a crucial step to help the Hough transform to better distinguish the linear segments at the lower end of the lamina. The influence of the steps taken in this part is illustrated in Fig. 6.

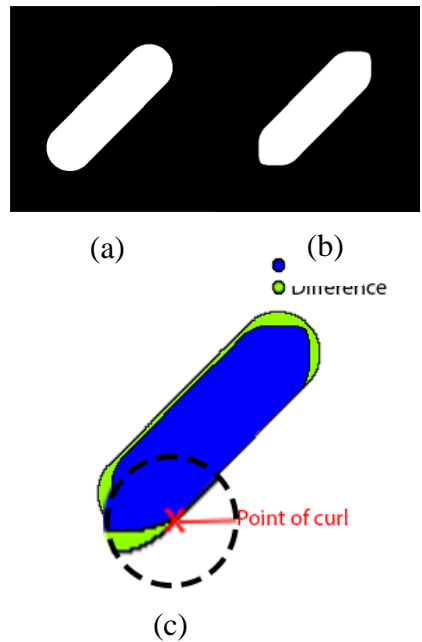

(a)  (b)

(c)

**Fig. 6** Impacts of the morphological steps on a synthetic lamina. a) Simple lamina representation b) two dilations with a square-shaped structuring element followed by an erosion with disk-shaped structuring element c) rectangular shape at the curved point of the lamina helps Hough Transform to distinguish the linear segments, more precisely.

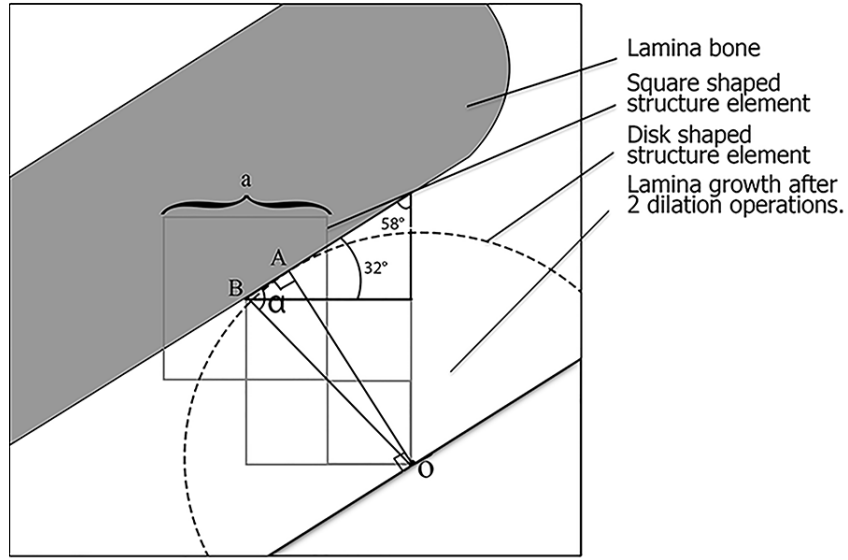

**Fig.7** Reversing the lamina shape from double dilated size to the eroded size.

If with a priori knowledge, we suppose the angle of the lamina is around $32°$[3], then from Fig. 7 the corresponding equation for the radius of the disk-shaped structuring element will be

$$\alpha = 77°, OA = a\sqrt{2} \times \sin(77°), \tag{7}$$

hence, the lamina will revert to its original spatial coordinations.

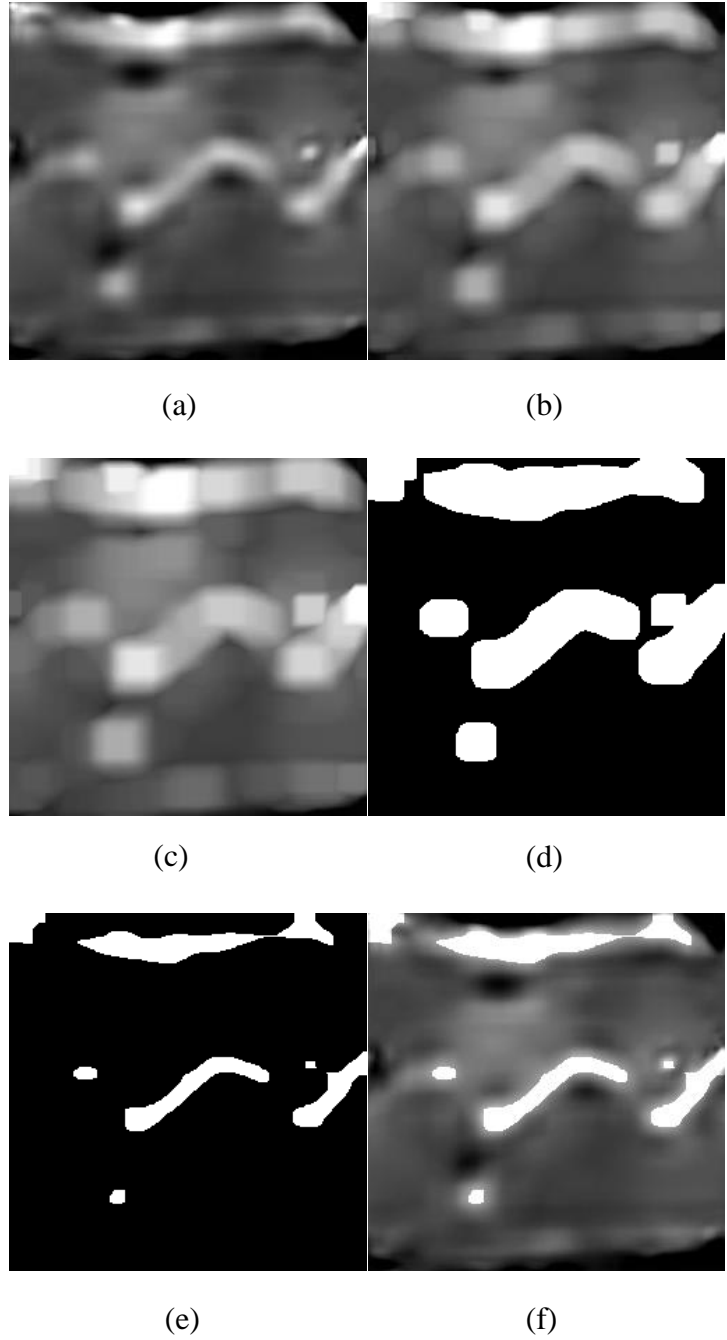

(a)          (b)

(c)          (d)

(e)          (f)

**Fig. 8** Intermediate images of our method: (a) the denoised image, (b) first dilation, (c) the second dilation, d) binarizing the image, e) erosion operator applied, f) overlay of the result on the input image.

The expressions for the lamina shape adjustment to match the prerequisites of the Hough Transform is

$$B = (A \oplus SE1) \oplus SE1,\ C = B \ominus SE2,\ D = threshold\{(A \oplus SE1) \oplus SE1\} \ominus SE2, \qquad (8)$$

where, ⊕ and ⊖ are the morphological dilation and erosion, respectively, $A$ is the denoised image, SE1 is an 8×8 pixels square-shaped structuring element, SE2 is a 10×10 pixels disk-shaped structuring element, and the *threshold* {} is the thresholding operator which binarizes the given image using the threshold value obtained from Otsu`s thresholding method multiplied by a constant value which is empirically chosen and is set to 0.9. After binarizing the image, a morphological erosion operator with a slightly larger structuring element obtained from Eq. 7 is employed on the binary image to recover the original size of the laminas. Next, a Prewitt edge detection filter which is a discrete filter operating in horizontal and vertical directions is chosen here to create the edge map required for the line detection steps by

$$G_x = \begin{bmatrix} +1 & 0 & -1 \\ +1 & 0 & -1 \\ +1 & 0 & -1 \end{bmatrix} * A, G_y = \begin{bmatrix} +1 & +1 & +1 \\ 0 & 0 & 0 \\ -1 & -1 & -1 \end{bmatrix} * A, \qquad (9)$$

where the gradient`s magnitude and direction are determined as follows:

$$|G| = \sqrt{G_x^2 + G_y^2}, \theta = tan^{-1}(G_y/G_x) \qquad (10)$$

Intermediate images of these operations are presented in Fig. 8. It is worth mentioning that the proportions of the structuring elements are held at the same values for all our datasets.

*C. Detection of linear pieces via Hough transform*

The Hough Transform (HT) is a fast mathematical operation to detect arbitrary shapes in an image. In our case, the linear shapes are of interest. The Hough Transform uses a parametric representation of a matrix whose rows and columns correspond to $\rho$ and $\theta$ values, respectively, and the elements in the HT matrix represent accumulator cells. It is the most reliable method to detect straight lines by mapping each active input pixel to accumulator space[30].

$$\rho = x\cos(\theta) + y\sin(\theta), \qquad (15)$$

where $\rho$ is the distance from the origin of the image to the line along a vector perpendicular to the line, $\theta$ is the angle between the x-axis of the image and the vector, and *(x,y)* is a point from the image plane.

The weights of the matrix will yield $\rho$ and $\theta$ values for each line. Due to imperfections in either the thresholded image or the edge map, there may be misplaced pixels on the desired lines as well as spatial deviations between a perfect line and the jagged lines in the edge map, thereby detecting a slightly curved line as two separate linear parts via Hough transform. To address this problem, the resolution of $\theta$ and $\rho$ parameters are set to experimentally chosen values to group the proper set of lines.

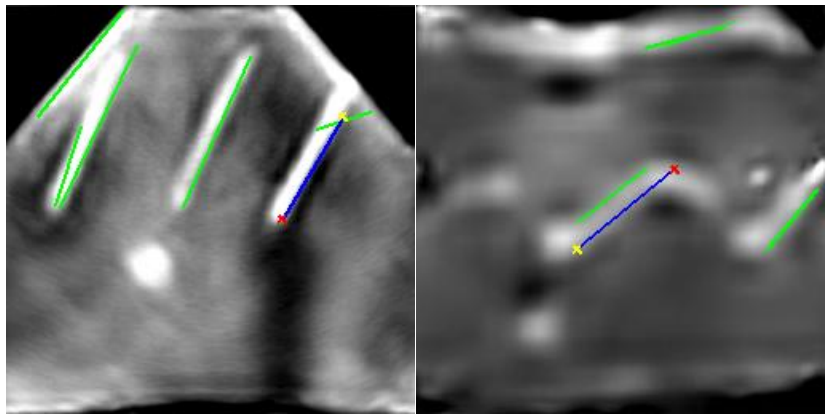

(a)                              (b)

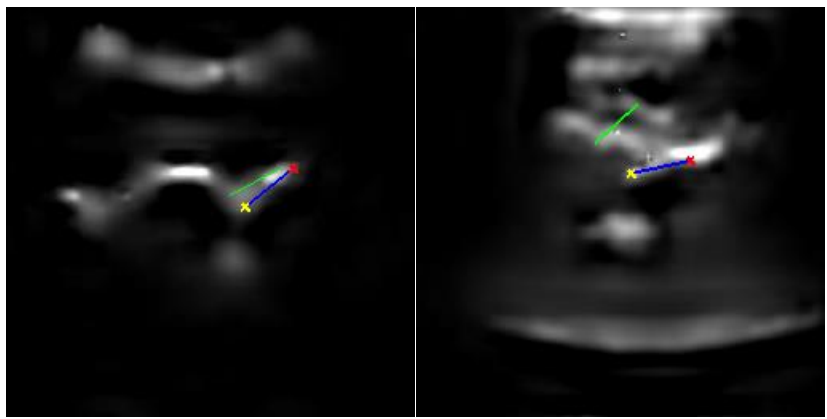

(c)                              (d)

**Fig. 9** Performance of the algorithm on: a) denoised synthetic image simulated via Filed II software. b) denoised image from[3]. c) another denoised image from our data set. As shown here the algorithm measures the epidural space for the laminas focused by the sonographer. d) ultrasound image of a broken lamina.

*3.2 Phase symmetry and thresholding as an alternative approach*

Phase symmetry is a well-known approach among ultrasound image processing techniques to highlight the bone structures, and is defined as Eq. 1, where $\varepsilon$ is a small value included to avoid division by zero, T is a threshold value to compensate for the noise energy response which equals to T= $\mu$ + k$\sigma$, where k is a user-specified number, and $\mu$ and $\sigma$ represent the mean and standard deviation of the Rayleigh distribution of the noise energy response, respectively. $e_n(x)$ and $o_n(x)$ are the even and odd parts of the response of Log-Gabor filter defined as

$$[e_n(x), o_n(x)] = [I(x) * M_n^e, I(x) * M_n^o], \tag{11}$$

where $I$ denotes the signal, and $M_n^e$ and $M_n^o$ denote the even and odd Log-Gabor filters at scale $n$.

$$G(\omega) = \exp\left(\frac{-(\log(\frac{\omega}{\omega_0}))^2}{2(\log(\frac{k}{\omega_0}))^2}\right), \tag{12}$$

$$M_n^e = real\left(F^{-1}(G(\omega))\right), \tag{13}$$

$$M_n^o = imag\left(F^{-1}(G(\omega))\right), \tag{14}$$

where $\frac{k}{\omega_0}$ is a ratio to be kept constant to achieve constant shape ratio filters, and $\omega_0$ is the filter`s center spatial frequency. $G(\omega)$ is the Log-Gabor filter and $F^{-1}$ is the inverse Fourier transform operation.

With the phase symmetry output in hand, two morphological dilation operations are employed on the phase symmetry output in a consecutive manner to account for the thin bone seed points resulted from phase symmetry application. The size of the structuring element is chosen empirically and is fixed at 3 pixels for all the data sets.

Finally, Otsu`s[40] thresholding method which minimizes the intra-class variances is applied in this step to provide a binary image of laminas.

With the use of phase symmetry on the denoised image, a better thresholding result is achieved. Fig. 5 shows the results of Otsu`s thresholding method both after and before employing phase symmetry.

The number of scales is 5, and the chosen orientations are 90,120,150. All log Gabor filters are multiplied by a low pass filter with a radius of 0.035 and a sharpness of 10. Empirically it seems that the noise effect is underestimated by a factor of 0.9 ($T_2=T/0.9$). The rest of the parameters were as discussed in an earlier study[29]. It is crucial to add the morphological step in this researched approach. As Fig. 9 shows, morphological operations help the method to recover the shape of the lamina and provide a better representation of its shape and a better estimation of epidural space depth, consequently.

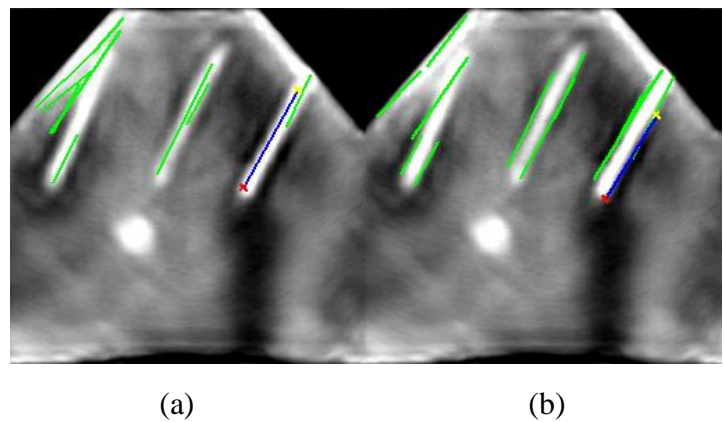

(a)          (b)

**Fig. 10** Alternative PS approach result on the synthetic image. a) phase symmetry without morphological operations. The parallel outlines of the laminas are fused together to form a single lamina shape disregarding the boundary. b) phase symmetry followed by morphological operations. The outer shape of the lamina is detected successfully.

*3.3 Contour Simplification using Ramer-Douglas-Peucker Algorithm:*

Contours ($Ci$) extracted from the edge map ($E$) often contain intricate details that may not be crucial for subsequent processing steps. In this context, we employ the Ramer-Douglas-Peucker (RDP) algorithm to simplify the contours, reducing the number of points required to represent the shapes while preserving essential features.

Given a contour $Ci$ with points $Pi1, Pi2, \ldots, Pin$, the RDP algorithm is defined as follows. The first and last points, Pi1 and Pin, form the initial line segment $Line(Pi1, Pin)$. For each intermediate point $P_{ij}$ ($2 \leq j \leq n-1$), the perpendicular distance $dist(P_{ij}, Line(P_{i1}, P_{in}))$ is computed. The point Pik with the maximum distance is identified ($Max_k\ dist(P_{ik}, Line(P_{i1}, P_{in}))$).

If $dist(P_{ik}, Line(P_{i1}, P_{in})) > \varepsilon$, the RDP algorithm is applied recursively to the sub-contours Pi1:k and Pik:n. Conversely, if $dist(P_{ik}, Line(P_{i1}, P_{in})) \leq \varepsilon$, the points between Pi1 and Pin are omitted in the simplified contour Ci.

The resulting simplified contour Ci retains the essential shape of the original contour with a reduced set of points, offering advantages in terms of computational efficiency and subsequent processing steps. This preprocessing step, particularly beneficial before applying methods like the Hough transform, ensures that the data representation focuses on salient structural elements while minimizing unnecessary detail.

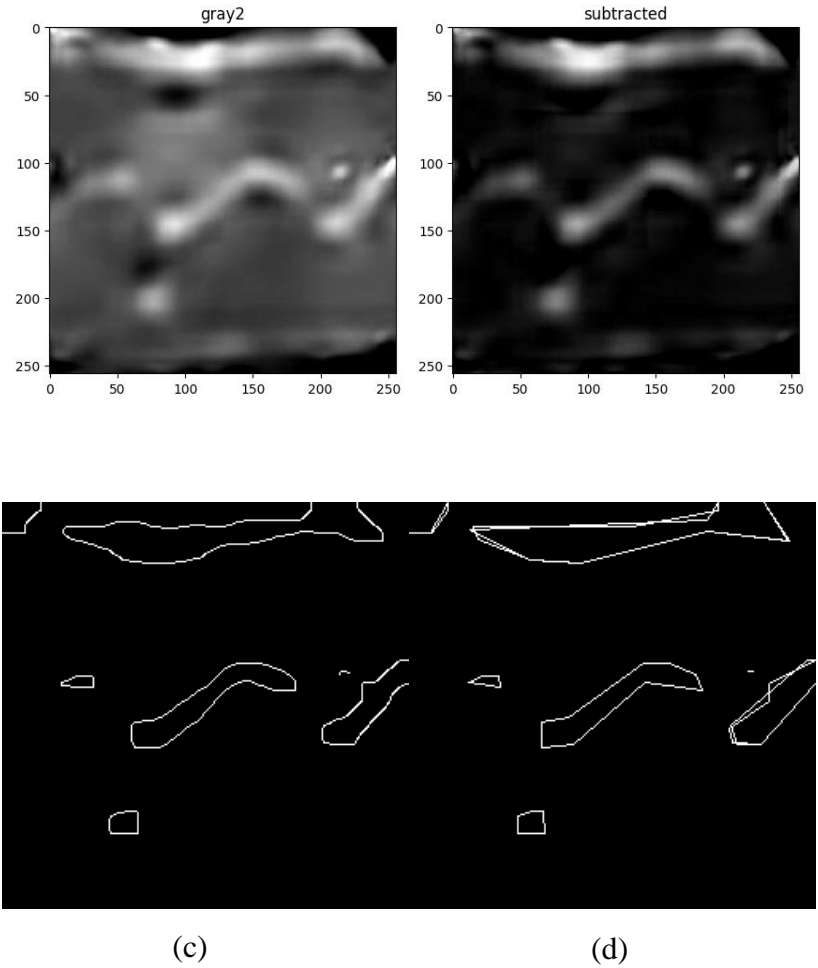

(c)　　　　　　　　　　　　(d)

**Fig. 11** Original Image. a) Subtracting the image with opening operation applied from the original image to enhance the lamina bone. b) Edge map. c) Contour simplification using Ramer-Douglas-Peucker-Algorithm. The corner of the lamina bone which was sharpened by the morphological steps is further enhanced in this step; even though, the ultimate purpose of this step is to increase the hough transform performance.

## 4   Clinical trials and methods

In this section, clinical trials and methods along with a fair evaluation of the proposed algorithm are presented

*4.1 Methods and evaluation*

On the whole, the performance of our method was analyzed on 16 ultrasound images, and a success rate of 93% was achieved. Looking into the failed detection case, the lamina was misdetected (see Fig. 11), because, that was a low-quality image of an obese patient with a relatively high BMI of 42 which was much larger than all other subjects, and the lamina was not discernable, thereby, failing the algorithm at thresholding step.

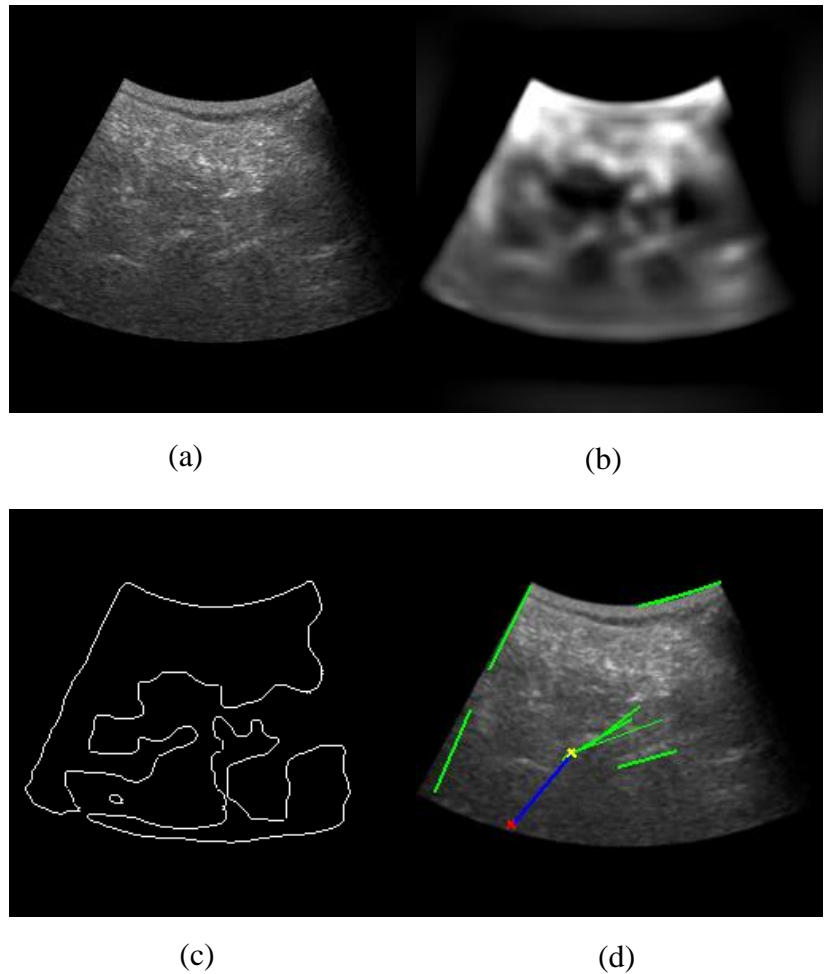

(a)              (b)

(c)              (d)

**Fig. 12** Failed detection of laminas. a) raw ultrasound image. b) denoised images using the CWD method. c) edge map of the thresholded image after morphological dilations. Unclear visibility of laminas prevented the thresholding step from differentiating between laminas and the homogeneous region of the image. d) result of the algorithm as an overlay on the image. Even more limited angle of search for the lamina could prevent the unrelated detection of segments.

Angles of the laminas are known to be around 32 °[3]. Therefore, the variable theta is limited to a range of 20° up to 75°. To avoid excessive detection of the straight lines by the Hough algorithm, the resolution of ρ and θ are set to 2 pixels and 5°, respectively, for all of our data sets. Fig. 12 shows the Hough Transform accumulator matrices and the corresponding laminas are detected as peaks at around an orientation of 40° which is close to the priori knowledge of almost 32°.

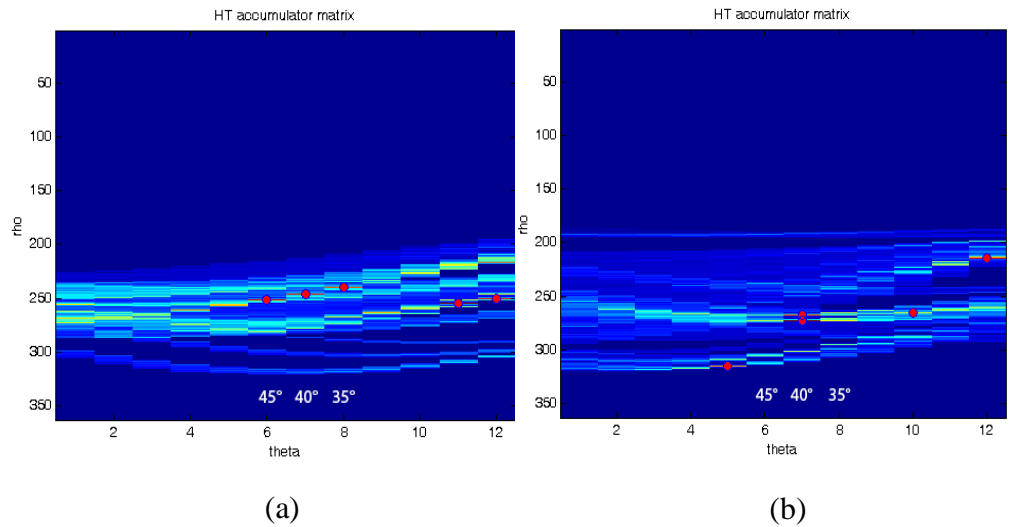

(a)  (b)

**Fig. 13** HT accumulator matrices for: (a) Fig. 9(b) and (b) Fig. 9(c).

The accuracy of the 10 successful detections of the epidural space is shown in Table 1. The RMS error of the automatic detection versus manual inspection of the sonographer is 0.07 mm, and 95% limits of agreement of -0.847 to 1.142 mm. The failed detection case is removed from our evaluations as we do not focus on parturient subjects in our study. The Bland-Altman plot is shown in Fig. 14.

**Tabel 1** The accuracy of the successful detections

| Automatic detection of proposed method versus | RMS error (mm) | Mean Absolute error(mm) | Bland-Altman 95% limits of agreement (mm) |
|---|---|---|---|
| Manual detection of sonographer | 0.07 | 0.62 | -0.847 to 1.142 |

The computational cost for each step of the proposed, alternatively analyzed, and LF detection methods are measured and shown in Table.2. The four primary operations of the proposed algorithm are the morphological dilations (0.26 s), thresholding and edge detection (0.4 s), and Hough Transform (0.11 s).

Even though five peaks of the hough accumulator matrix were enough to cover all lamina peaks, 20 line peaks on the Hough accumulator matrix were chosen for searching in the matrix to achieve a more robust algorithm (see Fig. 10(d))

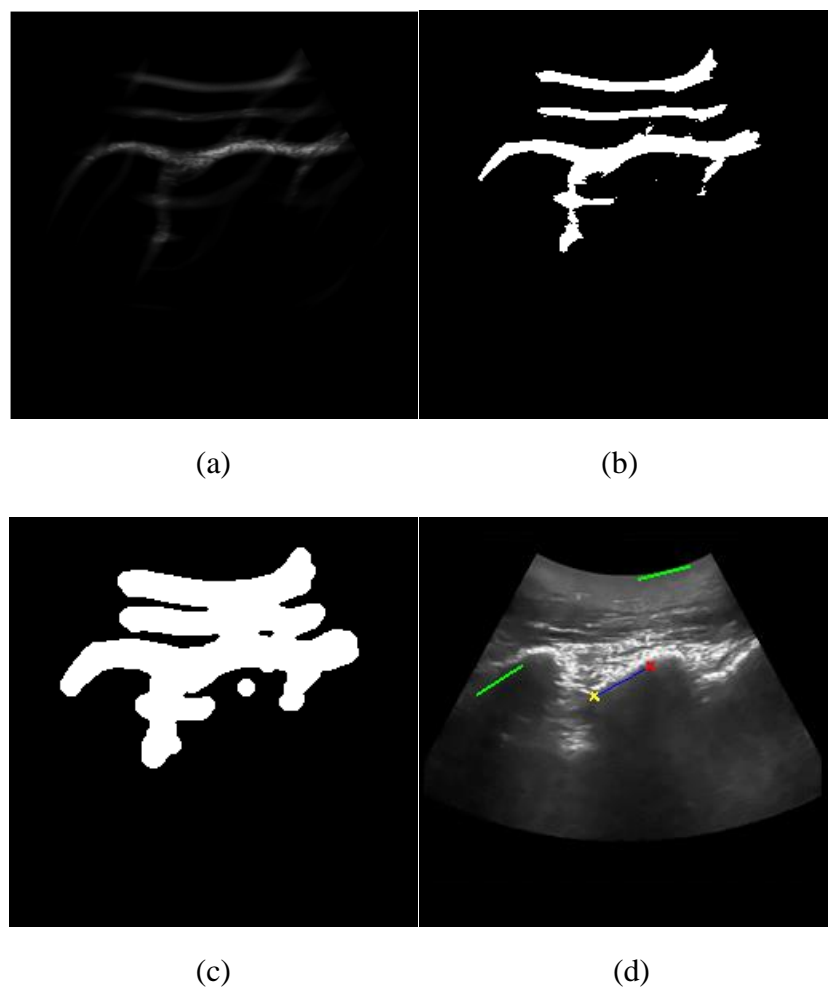

(a)            (b)

(c)            (d)

**Fig. 13** Analysis of the alternative approach using phase symmetry information. (a) phase symmetry of (d). (b) thresholded result of the method. (c) result of morphological operations of this analysis. (d) result of the alternative

method as an overlay which gives a good estimation of the epidural space depth, although the thresholding step shows that this approach could easily fail at detection as the excessive ridges are grown through morphological operations. Since the algorithm needs a clean image for line detection steps, Otsu`s threshold value is divided by a factor of 2 to remove unnecessary phase symmetry ridges from the image.

$$Threshold_{PS} = 0.5 \times Threshold_{Otsu} \qquad (16)$$

Fig. 13(b) shows that by using a manipulated thresholding value, phase symmetry ridges where the bone has no visibility has been removed.

The accuracy of the algorithm was measured manually by a visual review of the results. The depth of the epidural space was distinguished effectively if the identification was in line with the LF body.

So as to survey the precision of the algorithm, the difference between the determined distance from skin to epidural space and the manual detected depth was inspected. The average error, RMS error, and the Bland-Altman 95% limits of agreement were determined on the successfully distinguished epidural space depths. Despite the fact that the images can contain three laminas, just the calculation for the most central lamina was used since it is the one the sonographer focused on to highlight and along which the anesthesiologist inserted the needle.

Eventually, the time consumption of the algorithm was assessed. The algorithm was written in MATLAB™ and tested on 256×256 pixels images, comprising synthetic and real ultrasound images of lumbar anatomy in the paramedian plane. The synthetic image is shown in Fig. 13(b) which was created in Photoshop™ and then simulated in Field II™ software to project the ultrasound scanning device effects on the image. The code was run on an Intel Core i7 740QM at 1.73 GHz with 6-GB RAM.

**Table 2** Computational cost of Automatic detection of epidural space using MATLAB on an Intel Core 740QM at 1.73 GHz with 6-GB RAM on a 256 × 256 pixels image

| Function | Computational cost (s) of the proposed method | Computational cost (s) of the alternative method | Computational cost (s) of the LF detection method |
| --- | --- | --- | --- |

| Phase symmetry | - | 0.4 | 0.3 |
| --- | --- | --- | --- |
| Morphological dilations | 0.26 | 0.26 | - |
| Threshold and edge map | 0.4 | 0.4 | - |
| Hough Transform | 0.11 | 0.11 | - |
| Template matching | - | - | 3.2 |
| **Total** | **0.77** | **1.17** | **3.5** |

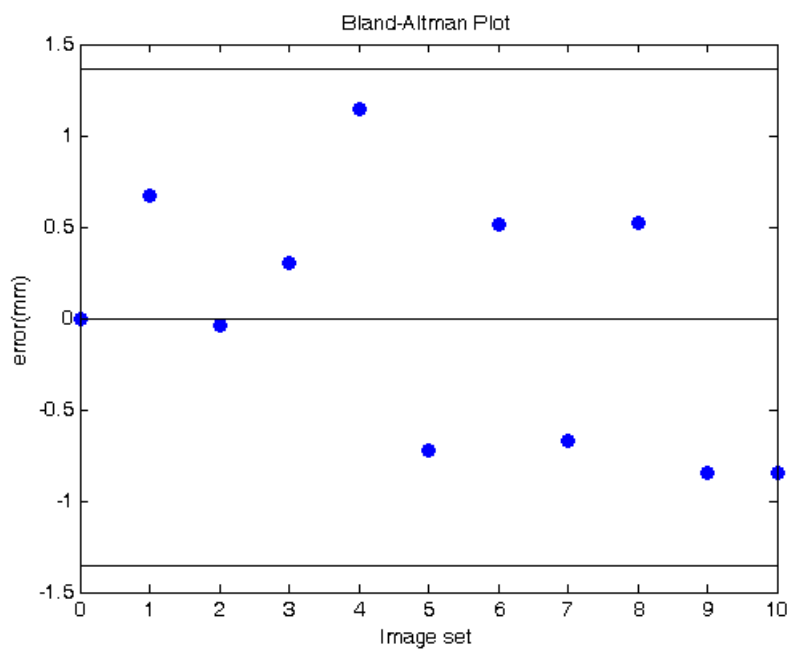

**Fig. 14** Bland Altman plot of automatic lamina detection versus medical expert measurements. The number of image sets n=10, the first five images are different varying intensity images, and the latter five are chosen from the second data set captured during the epidural needle insertions.

*4.2 Statistical significance*

The paired t-test was utilized to evaluate the statistical significance of the difference between the epidural space depths measured by the algorithm and the manual inspection of the image by the medical expert. For those assessments, the Pearson chi-square test was performed[20] to confirm the normality of the data.

*4.3 Clinical trial*

Ethical approval was obtained from the anesthesiology group of the Akhtar Educational Hospital to perform ultrasound scanning on patients scheduled for epidural or spinal anesthesia, and written consent was obtained from the participants.

Our data set contains 10 varying intensity ultrasound images, 5 of which were obtained from previous research by Sobhiyeh et al.[16] using a EUP-C314G ultrasound device with a 3.5 Mhz Curvilinear probe. For the other five images, scanning and data capture were performed by an experienced diagnostic medical sonographer (Dr. Haghighatkhah) using a SONOSITE EDGE II ultrasound device with a 2–5-MHz rC60xi Curvilinear transducer. The anesthesiologist (Dr. Mosaffa) performed all the epidural needle insertions, and the sonographer captured images for the offline processing and performed the measurements directly on the ultrasound image at a later time.

For further assessment, five image sequences were captured from each patient to analyze the algorithm in real-time situations. Even though the processing of images was done offline, promising results of the performance of the proposed automatic detection on image sequences illustrate that the proposed algorithm is a suitable approach for online image processing. See Table 1 for the time consumption measurements.

## 5  Conclusion and future works

The focus of the proposed method is on reporting the epidural space depth in a faster fashion. Hence, small misidentifications on the overall shape of the detected lamina are not taken into account for the epidural space depth measurements. Template matching approaches are still more reliable for exact epidural space localization, however, they suffer from high execution time in comparison with our method. In the future work, needle tip detection and prediction will be analyzed and the most possible location of the needle tip, especially in image sequences that needle tip discrepancy is not clear, will be reported based on the previous locations of the tip.

11. Grau, Thomas, et al. "Efficacy of ultrasound imaging in obstetric epidural anesthesia." *Journal of clinical anesthesia* 14.3 (2002): 169-175.

12. Pan, P. H., T. D. Bogard, and M. D. Owen. "Incidence and characteristics of failures in obstetric neuraxial analgesia and anesthesia: a retrospective analysis of 19,259 deliveries." *International journal of obstetric anesthesia* 13.4 (2004): 227-233.

13. Karmakar, M. K., et al. "Real-time ultrasound-guided paramedian epidural access: evaluation of a novel in-plane technique." *British journal of anaesthesia* 102.6 (2009): 845-854.

14. Hopkins, P. M. "Ultrasound guidance as a gold standard in regional anaesthesia." *British Journal of Anaesthesia* 98.3 (2007): 299-301.

15. Marhofer, P., M. Greher, and S. Kapral. "Ultrasound guidance in regional anaesthesia." *British Journal of Anaesthesia* 94.1 (2005): 7-17.

16. Chin, Ki Jinn, and Vincent Chan. "Ultrasound-guided peripheral nerve blockade." *Current Opinion in Anesthesiology* 21.5 (2008): 624-631.

17. Grau, T., et al. "Ultrasound imaging facilitates localization of the epidural space during combined spinal and epidural anesthesia." *Regional anesthesia and pain medicine* 26.1 (2001): 64-67.

18. Sobhiyeh, Sima, Amir Hossein Rezaie, and Hamid Reza Haghighatkhah. "Complex wavelet diffusion for enhancing ultrasound images of the laminar bone." *Journal of Electronic Imaging* 23.5 (2014): 053001-053001.

19. Tran, Denis, et al. "Preinsertion paramedian ultrasound guidance for epidural anesthesia." *Anesthesia & Analgesia* 109.2 (2009): 661-667.

20. Arzola, Cristian, et al. "Ultrasound using the transverse approach to the lumbar spine provides reliable landmarks for labor epidurals." *Anesthesia & Analgesia* 104.5 (2007): 1188-1192.

21. Jespersen, Søren K., Jens E. Wilhjelm, and Henrik Sillesen. "Multi-angle compound imaging." *Ultrasonic imaging* 20.2 (1998): 81-102.
32